\documentclass[twocolumn,prl,showpacs,preprintnumbers,amsmath,amssymb]{revtex4}
\usepackage{graphicx}
\usepackage{mathptmx,bm}
\begin{document}


\title{Fermi Surfaces and $p$-$d$ Hybridization in the Diluted Magnetic Semiconductor Ba$_{1-x}$K$_{x}$(Zn$_{1-y}$Mn$_{y}$)$_{2}$As$_{2}$ Studied by Soft X-ray Angle Resolved Photoemission Spectroscopy}

\author{H. Suzuki$^{1}$, G. Q. Zhao$^{2}$, K. Zhao$^{2}$, B. J. Chen$^{2}$, M. Horio$^{1}$, K. Koshiishi$^1$, J. Xu$^{1}$, M. Kobayashi$^{3}$, M. Minohara$^{3}$, E. Sakai$^{3}$, K. Horiba$^{3}$, H. Kumigashira$^{3}$, Bo Gu$^{4}$, S. Maekawa$^4$, Y. J. Uemura$^{5}$, C. Q. Jin$^{2,6}$ and A. Fujimori$^{1}$}
 
\affiliation{$^{1}$Department of Physics, University of Tokyo,
Bunkyo-ku, Tokyo 113-0033, Japan}

\affiliation{$^{2}$Beijing National Laboratory for Condensed Matter Physics, Institute of Physics, Chinese Academy of Sciences, Beijing 100190, China}

\affiliation{$^{3}$KEK, Photon Factory, Tsukuba, Ibaraki 305-0801, Japan}

\affiliation{$^{4}$Advanced Science Research Center, Japan Atomic Energy Agency, Tokai 319-1195, Japan}

\affiliation{$^{5}$Department of Physics, Columbia University, New York, New York 10027, USA}

\affiliation{$^{6}$Collaborative Innovation Center of Quantum Matter, Beijing, China}

\date{\today}

\begin{abstract}
The electronic structure of the new diluted magnetic semiconductor Ba$_{1-x}$K$_{x}$(Zn$_{1-y}$Mn$_{y}$)$_{2}$As$_{2}$ ($x=0.30$, $y=0.15$) in single crystal form has been investigated by angle-resolved photoemission spectroscopy (ARPES). 
Measurements with soft x-rays clarify the host valence-band electronic structure primarily composed of the As $4p$ states.  Two hole pockets around the $\Gamma$ point, a hole corrugated cylinder surrounding the $\Gamma$ and Z points, and an electron pocket around the Z point are observed, and explain the metallic transport of Ba$_{1-x}$K$_{x}$(Zn$_{1-y}$Mn$_{y}$)$_{2}$As$_{2}$. This is contrasted with Ga$_{1-x}$Mn$_{x}$As (GaMnAs), where it is located above the As $4p$ valence-band maximum (VBM) and no Fermi surfaces have been clearly identified. Resonance soft x-ray ARPES measurements reveal a nondispersive (Kondo resonance-like) Mn $3d$ impurity band near the Fermi level, as in the case of GaMnAs. However, the impurity band is located well below the VBM, unlike the impurity band  in GaMnAs, which is located around and above the VBM. 
We conclude that, while the strong hybridization between the Mn $3d$ and the As $4p$ orbitals plays an important role in creating the impurity band and inducing high temperature ferromagnetism in both systems, the metallic transport may predominantly occur in the host valence band in Ba$_{1-x}$K$_{x}$(Zn$_{1-y}$Mn$_{y}$)$_{2}$As$_{2}$ and in the impurity band in GaMnAs. 
\end{abstract}


\maketitle

After the discovery of ferromagnetism in Mn-doped GaAs, Ga$_{1-x}$Mn$_{x}$As (GaMnAs), diluted magnetic semiconductors (DMSs) have attracted considerable attention as a promising candidate for future spintronic devices  \cite{Ohno.H_etal.Science1998,Dietl.T_etal.Nat-Mater2010,Zutic.I_etal.Rev_mod_Phys2004,Ohno.H_etal.Applied-Physics-Letters1996,Dietl.T_etal.Science2000,Dietl.T_etal.Rev.-Mod.-Phys.2014}. While metal-based spintronic devices such as magnetoresistive random access memory are already developed to practical applications, semiconductor-based spin devices are yet to be explored. Considering the compatibility with existing semiconductor technologies, ferromagnetic (FM) semiconductors have a high potential. However, the prototypical material GaMnAs has the problems of limited chemical solubility of the magnetic element Mn and uncontrollable carrier density due to  defect formation. In order to realize desired functionalities of spintronic devices as well as to study the basic physics, it is desirable to have DMSs allowing for independent control of the carrier density and the concentration of magnetic ions. 

A novel series of DMS, Ba$_{1-x}$K$_{x}$(Zn$_{1-y}$Mn$_{y}$)$_{2}$As$_{2}$ (Mn-BaZn$_{2}$As$_{2}$), which is isostructural to the ``122''-type iron-based superconductors \cite{Paglione.J_etal.Nat-Phys2010}, was successfully synthesized with FM transition temperature ($T_{C}$) of 180 K \cite{Zhao.K_etal.Nat-Commun2013}, and recently the $T_{C}$ has risen up to $230$ K \cite{Zhao.K_etal.Chin.-Sci.-Bull.2014}, which are comparable or even higher than those of GaMnAs. This material has an advantage that the Ba layer, where hole carriers are generated, and the ZnAs layer, where magnetic elements are introduced, are spatially separated. This allows to independently control the concentration of charge carriers and magnetic elements. Also, the same chemical valence 2+ of  Mn and the host Zn atoms allows high chemical solubility of Mn, and makes it possible to obtain bulk specimens. Furthermore, interstitial Mn ions in the ZnAs layers are energetically unstable and thus Mn ions go only to substitutional sites \cite{Glasbrenner.J_etal.Phys.-Rev.-B2014}. This simplifies the theoretical treatment and experimental understanding of carrier-induced ferromagnetism in DMSs. In a previous work, we have performed an x-ray absorption (XAS) and resonance photoemission study of Mn-BaZn$_{2}$As$_{2}$ and found that the Mn $3d$ states are strongly hybridized with the ligand As $4p$ orbitals as in GaMnAs and that the Mn $3d$ partial density of states (PDOS) is similar to that of GaMnAs \cite{Suzuki.H_etal.Phys.-Rev.-B2015}.
 
Recently, Mn-BaZn$_{2}$As$_{2}$ bulk single crystals were successfully synthesized with $T_{C}$ = 60 K ($x=0.3$, $y=0.15$) \cite{unpub}. 
This is an important step toward the understanding of the electronic structure of this material and consequently the mechanism of carrier-induced ferromagnetism in DMSs in general. Also, a magnetic anisotropy has been found in the magnetization hysteresis curves with the easy axis along the $c$-axis \cite{unpub}. For future applications, this magnetic anisotropy leads to the possibility of high density storage device design.
In order to obtain more detailed information about the electronic structure and to discuss the similarities to and differences from GaMnAs, we have performed angle-resolved photoemission spectroscopy (ARPES) measurements on Mn-BaZn$_{2}$As$_{2}$ ($x=0.3$, $y=0.15$, $T_{C}$ = 60 K) single crystals. 

\begin{figure}[htbp] 
   \centering
   \includegraphics[width=9cm]{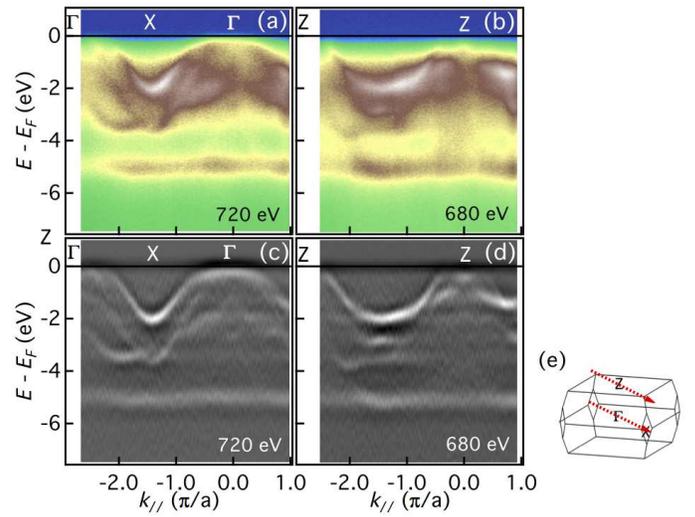} 
   \caption{(Color online) (a), (b) Soft x-ray angle-resolved photoemission spectroscopy (ARPES) intensity taken with $h\nu =$ 720 and 680 eV photons . (c), (d) Corresponding second derivatives with respect to energy. (e) The first Brillouin zone. The momentum cuts are shown by arrows.}
   \label{SX}
\end{figure}

Single crystals were grown 
by the arc-melting solid-state reaction method. 
XAS and ARPES experiments were performed at beamline 2A of Photon Factory, KEK. This beamline is equipped with two undulators for vacuum ultraviolet (VUV) and soft x-ray (SX) light and enables us to measure samples with a wide range of photons (from $\sim$ 20 eV to $\sim$1500 eV). The VUV and x-ray incident light was linearly $p$-polarized to the samples.  Samples were cleaved \textit{in situ} prior to the measurements to obtain fresh surfaces. 
XAS spectra were taken in the total-electron yield mode. The kinetic energies and momenta of photoelectrons were measured using a Scienta SES2002 electron analyzer. The total energy resolution was 
 20 meV and 150 meV for VUV and SX light, respectively. 
All the measurements were done at 20 K, which was well below $T_{C}$. Mn-BaZn$_{2}$As$_{2}$ has a space group symmetry I4/mmm and its first Brillouin zone (BZ) is shown in Fig. \ref{SX} (e). In-plane ($k_{X}$, $k_{Y}$) and out-of-plane momenta ($k_{z}$) are expressed in units of $\pi/a$ and $2\pi/c$, respectively, where $a=4.12$ \AA \,and $c=13.58$ \AA\, are the in-plane and out-of-plane lattice constants. The $X, Y$ axes point from Zn towards the second nearest Zn atoms and the $z$ axis is parallel to the $c$-axis.

   

A valence band ARPES spectrum taken with $h\nu$ = 60 eV for a cut containing the $\Gamma$ point is shown in Fig. S1 of the supplemental material \cite{supple}. Due to the low photoemission cross section of the As $4p$ orbitals at 60 eV \cite{Yeh.J_etal.At.-Data-Nucl.-Data-Tables1985} and the short probing depth of VUV photoemission, the As $4p$ valence band was difficult to detect, although the Zn $3d$ levels were clearly detected. Unlike the parent compound of iron-pnictide superconductors, BaFe$_{2}$As$_{2}$, where the Fe $3d$ orbitals make dominant contribution to the electronic states near the Fermi level ($E_{F}$), the Zn $3d$ level is located at $E \sim$  -10 eV as shown in the angle-integrated spectrum (Fig. S1 (b)), consistent with the previous SX photoemission result on polycrystals \cite{Suzuki.H_etal.Phys.-Rev.-B2015}.

For SX,  the photoemission cross section from the As $4p$ orbitals is larger and the probing depth is longer, providing us with information about the bulk electronic structure. Figures \ref{SX} (a) and (b) show ARPES spectra taken with $h\nu$ = 720 and 680 eV photons. Cuts in the first BZ are indicated in panel (e). 
The $k_{z}$ positions for these photon energies are approximately those of the $\Gamma$ and Z points, respectively. The corresponding second derivatives along the energy directions are also shown in Figs. \ref{SX} (c) and (d). We observe multiple dispersive bands near $E_{F}$.

\begin{figure}[htbp] 
   \centering
   \includegraphics[width=9cm]{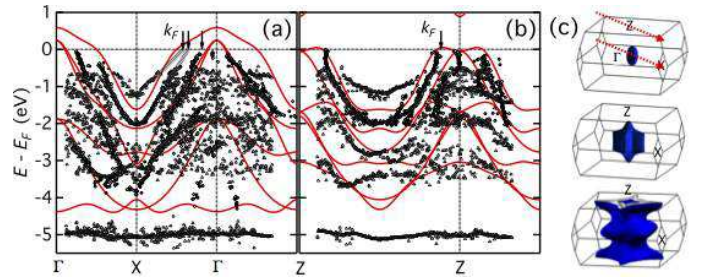} 
   \caption{(Color online) (a), (b) Comparison of the peak positions between the ARPES data and density-functional-theory band dispersion of the host material Ba$_{0.7}$K$_{0.3}$Zn$_{2}$As$_{2}$. Cuts are the same as those in Fig. \ref{SX}. The peak positions have been determined by fitting the energy distribution curves (triangles) and momentum distribution curves (circles) to Lorenzian functions convoluted with the instrumental Gaussian function. Arrows indicate experimental Fermi wave vectors ($k_{F}$'s) of the hole bands that form Fermi surfaces (FSs). The $k_{F}$'s of outer two bands in panel (a) is determined by linear extrapolation of the peaks to the Fermi level.  (c) Calculated FSs. Top: Hole FS, Middle: Hole FS, Bottom: Hole FS centered at the $\Gamma$ point and electron FS centered at the Z point.}
   \label{comp}
\end{figure}

To clarify the effect of hole and Mn doping on the electronic structure of the host semiconductor, we performed density-functional theory (DFT) band-structure calculation of the host semiconductor BaZn$_{2}$As$_{2}$ using the Wien2k package \cite{Blaha.P_etal.2001} and compare the result with the ARPES data. The calculations were done using experimentally determined tetragonal lattice constants $a=4.12$ \AA, $c=13.58$ \AA\ \cite{Zhao.K_etal.Nat-Commun2013} and the arsenic height $h_{\text{As}}=1.541$ \AA\ \cite{Hellmann.A_etal.Z.-Naturforsch.2007}.  First we calculated the band structure of BaZn$_{2}$As$_{2}$ and shfted the chemical potential downward so that the Luttinger volume matches the hole number provided by the K doping. For the exchange-correlation functional, 
we employed the modified Becke-Johnson exchange potential  
with the mixing factor for the exact-exchange term 
 of 0.25 \cite{Perdew.J_etal.The-Journal-of-Chemical-Physics1996}.

In Figs. \ref{comp} (a) and (b) we compare the peak positions of the ARPES data and the DFT band dispersion of Ba$_{0.7}$K$_{0.3}$Zn$_{2}$As$_{2}$. The peak positions have been determined by fitting the energy distribution curves (EDCs) and momentum distribution curves to Gaussian-convoluted Lorenzian functions. While the experimental dispersion is qualitatively well reproduced by the DFT calculation, the band bottom/top energies are not correctly predicted. Figure \ref{comp} (c) shows the calculated Fermi surfaces (FSs) of Ba$_{0.7}$K$_{0.3}$Zn$_{2}$As$_{2}$. There are two hole pockets around the $\Gamma$ point, a corrugated hole cylinder located at the zone center, and an electron pocket around the Z point. The three hole-like dispersions around the $\Gamma$ point and hole-like dispersion around the Z point observed by ARPES can be assigned to the calculated three hole FSs. The Fermi momenta ($k_{F}$'s) in units of $\pi/a$ are 0.6 (outer), 0.50 (middle) and 0.26 (inner) around the $\Gamma$ point, and 0.33 around the $Z$ point, respectively, as shown by the arrows in Fig. \ref{comp} (a) and (b). $k_{F}$'s in the outer two bands around the $\Gamma$ point are estimated by linear extrapolation of the band dispersions.  While we cannot clearly resolve the tiny electron pocket around the Z point, the slightly enhanced intensity at $E=0$, $k_{\parallel}=0$ suggests the presence of an electron pocket. The presence of FSs in the host material is contrasted with GaMnAs, where no clear Fermi surface crossing has been observed \cite{Kobayashi.M_etal.Phys.-Rev.-B2014}.


\begin{figure}[htbp] 
   \centering
   \includegraphics[width=5cm]{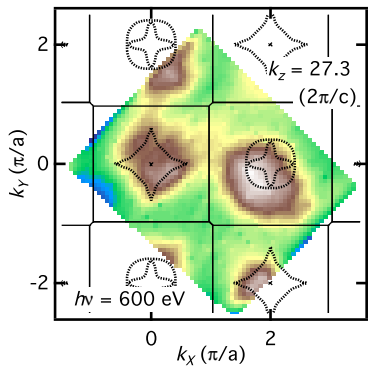} 
   \caption{(Color online) Photoemission intensity integrated within the energy window of $\Delta E=\pm 250$ meV. Dotted lines show the calculated FSs.}
   \label{mapping}
\end{figure}

Constant energy mapping at $E_{F}$ is shown in Fig. \ref{mapping}. To obtain the mapping, ARPES intensity within the energy window of $\Delta E=\pm 250$ meV has been integrated. Strong intensities appear around (0, 0) and (2$\pi$, 0), consistent with the calculated FSs located at the zone center. 

\begin{figure}[htbp] 
   \centering
   \includegraphics[width=9cm]{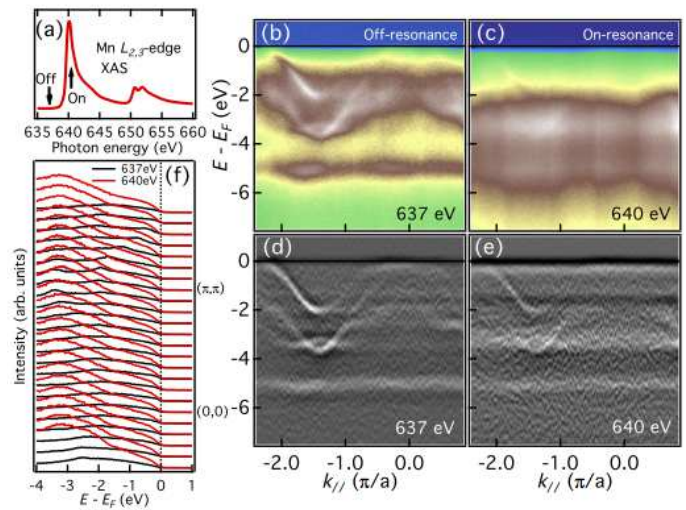} 
   \caption{(Color online) (a) Mn $L_{2,3}$-edge x-ray absorption spectrum of Mn-BaZn$_{2}$As$_{2}$. Off- and on-resonance photon energies are chosen to be 637 eV and 640 eV, respectively. (b), (c) ARPES energy-momentum intensity taken with on- and off-resonance energy photons. (e), (f) Second derivatives of the on- and off-resonance ARPES spectra. (f) Energy distribution curves of on- (red) and off- (black) resonance ARPES spectra.}
  
   \label{RPES}
\end{figure}

In order to extract information about the Mn $3d$ impurity levels, we performed resonance ARPES measurements around the Mn $L_{3}$-edge. At the $L_{3}$ edge, the direct photoemission of a 3$d$ electron positively interferes with the absorption followed by a Coster-Kr$\ddot{\text{o}}$nig transition  \cite{Gelmukhanov.F_etal.Physics-Reports1999,Bruhwiler.P_etal.Rev.-Mod.-Phys.2002}. This can be utilized to selectively enhance the photoemission intensity from the $3d$ levels of a transition metal in solids. From the XAS spectra shown in Fig. \ref{RPES} (a), we determined the on- and off-resonance energies to be 637 eV and 640 eV, respectively. ARPES spectra taken with these energies are shown in Figs. \ref{RPES} (b) and (c). The change in these spectra enables us to clearly identify the resonance enhancement of Mn $3d$-related photoemission features around -3.5 eV. From the color plot we observe enhanced nondispersive intensity due to the Mn $3d$ states overlaid on the the host bands, which are barely seen due to the overwhelmingly strong Mn $3d$-related features. The strong enhancement around -3.5 eV corresponds to the main peak in the Mn $3d$ PDOS  \cite{Suzuki.H_etal.Phys.-Rev.-B2015}.
  
In order to further clarify the Mn $3d$-derived states near $E_{F}$, we plot in Figs. \ref{RPES} (d) and (e) the second derivatives of the on- and off-resonance ARPES spectra and compare the corresponding EDCs in Fig. \ref{RPES} (f). By comparing panels (d) and (e), one finds that a nondispersive feature appears slightly below $E_{F}$ ($\sim 0.3$ eV) for all momenta. The EDCs taken on resonance (Fig. \ref{RPES} (f)) proves that the near-$E_{F}$ feature does not originate from the Fermi cutoff but from non-dispersive states located slightly below $E_{F}$. A dispersionless ``impurity band'' near $E_{F}$ has also been found in a resonance SX ARPES study of GaMnAs \cite{Kobayashi.M_etal.Phys.-Rev.-B2014}, where Mn $3d$-derived impurity band hybridized with the light-hole band of the host GaAs has been identified. Note that the impurity band is located deep in the valence bands in Mn-BaZn$_{2}$As$_{2}$, while it is located near the valence-band maximum (VBM) and is almost split off from it in GaMnAs. The appearance of the Mn $3d$ PDOS near $E_{F}$ in addition to the main peak away from $E_{F}$ is reminiscent of the Kondo resonance observed in Ce compounds \cite{Allen.J_etal.Phys.-Rev.-B1983}. 

Now we discuss possible origins of the impurity bands in the Mn-doped DMSs. While Mn-BaZn$_{2}$As$_{2}$ and GaMnAs are shown to have ``impurity bands'',  it has been reported that impurity band is absent in another $p$-type DMS In$_{1-x}$Mn$_{x}$As (InMnAs) \cite{Okabayashi.J_etal.Phys.-Rev.-B2002}. 
 The highest $T_{C}$'s so far reported are 230 K for Mn-BaZn$_{2}$As$_{2}$ \cite{Zhao.K_etal.Chin.-Sci.-Bull.2014}, 185 K for GaMnAs \cite{Wang.M_etal.Appl.-Phys.-Lett.2008}, and 55 K for InMnAs \cite{Slupinski.T_etal.J.-Cryst.-Growth2002}. In InMnAs, free carriers rather than carriers in the impurity band explain the existence of the Drude component in the optical conductivity \cite{Hirakawa.K_etal.Phys.-Rev.-B2002,Hirakawa.K_etal.Physica-E2001}. In Ref. \onlinecite{Okabayashi.J_etal.Phys.-Rev.-B2002}, the origin of the absence of impurity band is attributed to the weaker hybridization between the Mn $3d$ orbitals and the $4p$ orbitals of As ligands due to longer Mn-As distance than GaMnAs. 
The present observation of the impurity band in Mn-BaZn$_{2}$As$_{2}$ corroborates the importance of the strong hybridization for the emergence of impurity levels and the high-$T_{C}$ carrier-induced ferromagnetism.
Note that the metallic transport in Mn-BaZn$_{2}$As$_{2}$ is caused by the multiple host bands crossing $E_{F}$, whereas the VBM of GaMnAs is located below $E_{F}$ and the Mn $3d$ impurity band is responsible for metallic transport in GaMnAs. These observations suggest that, in spite of the different nature of carriers which mediate magnetic interactions between the Mn spins, ferromagnetism is induced as long as the Mn $3d$-As $4p$ hybridization is strong enough. 

Finally, we discuss the relationship between the present results and the recent theoretical study by Glasbrenner \textit{et al.} \cite{Glasbrenner.J_etal.Phys.-Rev.-B2014}. First, the supercell calculation in Ref. \onlinecite{Glasbrenner.J_etal.Phys.-Rev.-B2014} shows that the main Mn $3d$ PDOS in Ba(Zn$_{0.875}$Mn$_{0.125}$)$_{2}$As$_{2}$ is located at $\sim$ -3.4 eV and its width is less than 1 eV. While the peak position is consistent with our observations, the width is underestimated as compared with the wide Mn $3d$ PDOS \cite{Suzuki.H_etal.Phys.-Rev.-B2015}. Second, the calculation predicts the emergence of an $\uparrow$ band as a result of hybridization between the As $4p$ and the narrow Mn $3d$ band and the $\uparrow$ band disperses from the original Mn $3d \uparrow$ level toward $E_{F}$ and above. However, in our experiment, the Mn $3d$-related spectral weight is broad and nondispersive up to $E_{F}$. This may reflect  the strong local electron correlations among the Mn $3d$ electrons and the randomness of the Mn atoms substituting the Zn sites. As for the reduction of the saturated magnetic moment of Mn from 5 $\mu_{B}$ expected from the high-spin configuration of Mn $3d$ electrons to $\sim$1 $\mu_{B}$, it can be explained by the competition between the antiferromagnetic (AFM) superexchange coupling between the nearest-neighbor Mn spins and the longer-range FM coupling between them mediated by As holes as described in Ref. \onlinecite{Glasbrenner.J_etal.Phys.-Rev.-B2014}. Indeed, the end material BaMn$_{2}$As$_{2}$ is an AFM insulator with a high N\'eel temperature ($T_{N}$) of 625 K and the hole-doped metallic system Ba$_{1-x}$K$_{x}$Mn$_{2}$As$_{2}$ remains AFM with a slight reduction of $T_{N}$ (480 K for $x$ $\sim$ 0.4) \cite{Pandey.A_etal.Phys.-Rev.-Lett.2012}. In order to achieve ferromagnetism, therefore, the average Mn-Mn distance must be sufficiently elongated and the AFM interaction strength must be reduced as in the present dilute Mn case. Interestingly, it has been reported that the As $4p$ conduction band in Ba$_{1-x}$K$_{x}$Mn$_{2}$As$_{2}$ shows itinerant ferromagnetism and coexists with the Mn AFM order \cite{Pandey.A_etal.Phys.-Rev.-Lett.2013,Ueland.B_etal.Phys.-Rev.-Lett.2015}. It is, therefore, natually expected that such an unconventional magnetic state, in which FM and AFM interactions coexist and compete with each other, can exist in Mn-BaZn$_{2}$As$_{2}$ as well. Direct investigation of the spin configurations in the magnetically ordered state of Mn-BaZn$_{2}$As$_{2}$ is desired in future studies.

In conclusion, we have studied the electronic structure of Ba$_{1-x}$K$_{x}$(Zn$_{1-y}$Mn$_{y}$)$_{2}$As$_{2}$ by SX ARPES. We observed the valence bands of the BaZn$_{2}$As$_{2}$ host composed of As $4p$ states. Two hole pockets around the $\Gamma$ point, a hole corrugated cylinder containing the $\Gamma$ point, and an electron pocket around the Z point  give rise to the conduction in Mn-BaZn$_{2}$As$_{2}$. Resonance SX ARPES results suggest the presence of Mn $3d$ impurity band near $E_{F}$ as in GaMnAs. We propose that the strong hybridization between the Mn $3d$ and the As $4p$ states plays an important role in creating the impurity band and inducing high temperature ferromagnetism, but that the metallic transport can be induced either by the host valence band or the impurity band in Mn-doping based DMSs. 


This work was supported by a Grant-in-Aid for Scientific Research from the JSPS (S22224005,15H02109), the MEXT Element Strategy Initiative to Form Core Research Center, Reimei Project from Japan Atomic Energy Agency, 
Friends of Todai Foundation in New York, and the US NSF PIRE (Partnership for International Research and Education: OISE-0968226). Work at IOPCAS was supported by NSF \& MOST as well as CAS International Cooperation Program of China through Research Projects. Experiment at Photon Factory was approved by the Photon Factory Program Advisory Committee (Proposal Nos. 2013S2-002, 2014G177). H.S. acknowledges financial support from Advanced Leading Graduate Course for Photon Science (ALPS) and the JSPS Research Fellowship for Young Scientists. 

\bibliography{SXARPES_DMS}

\begin{thebibliography}{10}

\bibitem{Ohno.H_etal.Science1998}
H. Ohno, Science {\bf 281},  951  (1998).

\bibitem{Dietl.T_etal.Nat-Mater2010}
T. Dietl, Nat. Mater. {\bf 9},  965  (2010).

\bibitem{Zutic.I_etal.Rev_mod_Phys2004}
I. \ifmmode \check{Z}\else \v{Z}\fi{}uti\ifmmode~\acute{c}\else \'{c}\fi{}, J.
  Fabian, and S. Das~Sarma, Rev. Mod. Phys. {\bf 76},  323  (2004).

\bibitem{Ohno.H_etal.Applied-Physics-Letters1996}
H. Ohno, A. Shen, F. Matsukura, A. Oiwa, A. Endo, S. Katsumoto, and Y. Iye,
  Appl. Phys. Lett. {\bf 69},  363  (1996).

\bibitem{Dietl.T_etal.Science2000}
T. Dietl, H. Ohno, F. Matsukura, J. Cibert, and D. Ferrand, Science {\bf 287},
  1019  (2000).

\bibitem{Dietl.T_etal.Rev.-Mod.-Phys.2014}
T. Dietl and H. Ohno, Rev. Mod. Phys. {\bf 86},  187  (2014).

\bibitem{Paglione.J_etal.Nat-Phys2010}
J. Paglione and R.~L. Greene, Nat. Phys. {\bf 6},  645  (2010).

\bibitem{Zhao.K_etal.Nat-Commun2013}
K. Zhao, Z. Deng, X.~C. Wang, W. Han, J.~L. Zhu, X. Li, Q.~Q. Liu, R.~C. Yu, T.
  Goko, B. Frandsen, L. Liu, F. Ning, Y.~J. Uemura, H. Dabkowska, G.~M. Luke,
  H. Luetkens, E. Morenzoni, S.~R. Dunsiger, A. Senyshyn, P. Bﾃｶni, and
  C.~Q. Jin, Nat. Commun. {\bf 4},  1442  (2013).

\bibitem{Zhao.K_etal.Chin.-Sci.-Bull.2014}
K. Zhao, B. Chen, G. Zhao, Z. Yuan, Q. Liu, Z. Deng, J. Zhu, and C. Jin, Chin.
  Sci. Bull. {\bf 59},  2524  (2014).

\bibitem{Glasbrenner.J_etal.Phys.-Rev.-B2014}
J.~K. Glasbrenner, I. \ifmmode \check{Z}\else
  \v{Z}\fi{}uti\ifmmode~\acute{c}\else \'{c}\fi{}, and I.~I. Mazin, Phys. Rev.
  B {\bf 90},  140403  (2014).

\bibitem{Suzuki.H_etal.Phys.-Rev.-B2015}
H. Suzuki, K. Zhao, G. Shibata, Y. Takahashi, S. Sakamoto, K. Yoshimatsu, B.~J.
  Chen, H. Kumigashira, F.-H. Chang, H.-J. Lin, D.~J. Huang, C.~T. Chen, B. Gu,
  S. Maekawa, Y.~J. Uemura, C.~Q. Jin, and A. Fujimori, Phys. Rev. B {\bf 91},
  140401  (2015).

\bibitem{unpub}
G. Q. Zhao {\sl et al}., (unpublished).

\bibitem{supple}
{S}ee Supplemental Material at @to be added@ for ARPES spectrum taken with
  $h\nu$ = 60 eV light.

\bibitem{Yeh.J_etal.At.-Data-Nucl.-Data-Tables1985}
J.-J. Yeh and I. Lindau, At. Data Nucl. Data Tables {\bf 32},    (1985).

\bibitem{Blaha.P_etal.2001}
P. Blaha, K. Schwarz, G.~K.~H. Madsen, D. Kvasnicka, and J. Luitz, {\em
  {WIEN2K}, {A}n {A}ugmented {P}lane {W}ave + {L}ocal {O}rbitals {P}rogram for
  {C}alculating {C}rystal {P}roperties} (Technische Universit\"{a}t Wien, Wien,
  Austria, 2001).

\bibitem{Hellmann.A_etal.Z.-Naturforsch.2007}
A. Hellmann, A. L{\"o}hken, A. Wurth, and A. Mewis, Z. Naturforsch. {\bf 62b},
    (2007).

\bibitem{Perdew.J_etal.The-Journal-of-Chemical-Physics1996}
J.~P. Perdew, M. Ernzerhof, and K. Burke, J. Chem. Phys. {\bf 105},  9982
  (1996).

\bibitem{Kobayashi.M_etal.Phys.-Rev.-B2014}
M. Kobayashi, I. Muneta, Y. Takeda, Y. Harada, A. Fujimori, J. Krempask\'y, T.
  Schmitt, S. Ohya, M. Tanaka, M. Oshima, and V.~N. Strocov, Phys. Rev. B {\bf
  89},  205204  (2014).

\bibitem{Gelmukhanov.F_etal.Physics-Reports1999}
F. Gel'mukhanov and H. {\AA}gren, Phys. Rep. {\bf 312},  87  (1999).

\bibitem{Bruhwiler.P_etal.Rev.-Mod.-Phys.2002}
P.~A. Br\"uhwiler, O. Karis, and N. M\aa{}rtensson, Rev. Mod. Phys. {\bf 74},
  703  (2002).

\bibitem{Allen.J_etal.Phys.-Rev.-B1983}
J.~W. Allen, S.~J. Oh, M.~B. Maple, and M.~S. Torikachvili, Phys. Rev. B {\bf
  28},  5347  (1983).

\bibitem{Okabayashi.J_etal.Phys.-Rev.-B2002}
J. Okabayashi, T. Mizokawa, D.~D. Sarma, A. Fujimori, T. Slupinski, A. Oiwa,
  and H. Munekata, Phys. Rev. B {\bf 65},  161203  (2002).

\bibitem{Wang.M_etal.Appl.-Phys.-Lett.2008}
M. Wang, R.~P. Campion, A.~W. Rushforth, K.~W. Edmonds, C.~T. Foxon, and B.~L.
  Gallagher, Appl. Phys. Lett. {\bf 93},    (2008).

\bibitem{Slupinski.T_etal.J.-Cryst.-Growth2002}
T. Slupinski, H. Munekata, and A. Oiwa, J. Cryst. Growth {\bf 237},  1331
  (2002).

\bibitem{Hirakawa.K_etal.Phys.-Rev.-B2002}
K. Hirakawa, S. Katsumoto, T. Hayashi, Y. Hashimoto, and Y. Iye, Phys. Rev. B
  {\bf 65},  193312  (2002).

\bibitem{Hirakawa.K_etal.Physica-E2001}
K. Hirakawa, A. Oiwa, and H. Munekata, Physica E {\bf 10},  215  (2001).

\bibitem{Pandey.A_etal.Phys.-Rev.-Lett.2012}
A. Pandey, R.~S. Dhaka, J. Lamsal, Y. Lee, V.~K. Anand, A. Kreyssig, T.~W.
  Heitmann, R.~J. McQueeney, A.~I. Goldman, B.~N. Harmon, A. Kaminski, and
  D.~C. Johnston, Phys. Rev. Lett. {\bf 108},  087005  (2012).

\bibitem{Pandey.A_etal.Phys.-Rev.-Lett.2013}
A. Pandey, B.~G. Ueland, S. Yeninas, A. Kreyssig, A. Sapkota, Y. Zhao, J.~S.
  Helton, J.~W. Lynn, R.~J. McQueeney, Y. Furukawa, A.~I. Goldman, and D.~C.
  Johnston, Phys. Rev. Lett. {\bf 111},  047001  (2013).

\bibitem{Ueland.B_etal.Phys.-Rev.-Lett.2015}
B.~G. Ueland, A. Pandey, Y. Lee, A. Sapkota, Y. Choi, D. Haskel, R.~A.
  Rosenberg, J.~C. Lang, B.~N. Harmon, D.~C. Johnston, A. Kreyssig, and A.~I.
  Goldman, Phys. Rev. Lett. {\bf 114},  217001  (2015).

\end{thebibliography}

\end{document}